\documentclass[useAMS,usenatbib,usegraphicx]{mn2e}

\usepackage{graphicx}
\usepackage[latin1]{inputenc}
\usepackage{color}
\usepackage{times}
\usepackage{natbib}
\usepackage{setspace}
\newif\ifAMStwofonts
\AMStwofontstrue
\definecolor{red}{rgb}{1,0.,0.}

\newcommand{\munich}{{\sc l-galaxies }}
\newcommand{\lcdm}{$\Lambda$CDM }

\def\lesssim{\lower.5ex\hbox{$\; \buildrel < \over \sim \;$}}
\def\gtrsim{\lower.5ex\hbox{$\; \buildrel > \over \sim \;$}}
\title[Galaxy formation and early dark energy] {Semi-analytic galaxy
  formation in early dark energy cosmologies}

\author[Fontanot et al.]{
  \parbox[t]{\textwidth}{Fabio Fontanot$^{1,2}$\thanks{E-mail:
      fabio.fontanot@h-its.org}, Volker Springel$^{1,3}$, Raul
    E. Angulo$^{4}$, Bruno Henriques$^{4}$}
    \vspace*{8pt}\\
    $^1$ Heidelberger Institut f\"ur Theoretische Studien (HITS), Schloss-Wolfsbrunnenweg 35, 69118 Heidelberg, Germany \\
    $^2$ Institut f\"ur Theoretische Physik, Philosophenweg 16, 69120 Heidelberg, Germany \\
    $^3$ Zentrum f\"ur Astronomie der Universit\"at Heidelberg, ARI, M\"onchhofstrasse 12-14, 69120 Heidelberg, Germany \\
    $4$ Max-Planck-Institute for Astrophysics, Karl-Schwarzschild-Str. 1, 85740 Garching, Germany}

\begin{document}
\date{Accepted ... Received ...}

\maketitle

\begin{abstract} 
  We study the impact of early dark energy (EDE) cosmologies on galaxy
  properties by coupling high-resolution numerical simulations with
  semi-analytic modeling (SAM) of galaxy formation and evolution. EDE
  models are characterized by a non-vanishing high-redshift
  contribution of dark energy, producing an earlier growth of
  structures and a modification of large-scale structure evolution.
  They can be viewed as typical representatives of non-standard dark
  energy models in which only the expansion history is modified, and
  hence the impact on galaxy formation is indirect.  We show that in
  EDE cosmologies the predicted space density of galaxies is enhanced
  at all scales with respect to the standard \lcdm scenario, and the
  corresponding cosmic star formation history and stellar mass density
  is increased at high-redshift. We compare these results with a set
  of theoretical predictions obtained with alternative SAMs applied to
  our reference \lcdm simulation, yielding a rough measure of the
  systematic uncertainty of the models.  We find that the
  modifications in galaxy properties induced by EDE cosmologies are of
  the same order of magnitude as intra-SAM variations for a standard
  \lcdm realization (unless rather extreme EDE models are considered),
  suggesting that is difficult to use such predictions alone to
  disentangle between different cosmological scenarios. However, when
  independent information on the underlying properties of host dark
  matter haloes is included, the SAM predictions on galaxy bias may
  provide important clues on the expansion history and the
  equation-of-state evolution.
\end{abstract}


\section{Introduction}\label{sec:intro}
The last decade has seen considerably advances in our understanding of
the properties of our Universe as a whole, in particular thanks to the
accurate measurement of the most important cosmological parameters
(see e.g. \citealt{Komatsu09} and references herein). One of the most
surprising discoveries of this ``precision cosmology'' epoch is that
some unknown form of {\it Dark Energy} (DE, hereafter) accounts for
more than $70\%$ of the energy density of the present-day Universe,
and is responsible for its accelerated expansion today (see e.g.,
\citealt{Perlmutter99}). The physical nature of DE, together with its
origin and time evolution, is one of the most enigmatic puzzles in
modern cosmology.

In the standard \lcdm cosmological model, DE is treated as a classic
cosmological constant (following Einstein's original conjecture) where
a homogeneous and static energy density fills the whole Universe. This
simple assumption, however, gives rise to a number of theoretical
problems, such as the high degree of ``fine-tuning'' required to
accommodate the present day value of $\Omega_\Lambda$ (see
\citealt{Weinberg89} for a review). For this reason, many alternative
scenarios to explain DE and the accelerated expansion have been
proposed, ranging from scalar field models such as quintessence to
radical modifications of the laws of gravity. Most observational
efforts currently concentrate on constraining effective
parametrisations of the DE equation of state (see
e.g. \citealt{Wetterich88, RatraPeebles88}), which is adequate for
`non-coupled' dark energy models in which the dark energy can be
approximated as uniform and influences structure formation only
through a modification of the cosmic expansion rate.

An interesting sub-class of these models are scenarios that involve a
non-negligible DE contribution at early times during recombination and
primordial structure formation, which also implies non-negligible
modifications of the cosmic microwave background \citep{Doran01}, of
big-bang nucleosynthesis \citep{Muller04} and of large-scale structure
formation \citep{Bartelmann06}. The non-linear structure formation
predicted in such early dark energy (EDE) models has been studied
through high-resolution $N$-body simulation \citep{Baldi12,
  GrossiSpringel09}. In particular, the impact on the statistical
properties of dark matter (DM) haloes (such as the abundance of
high-redshift galaxy clusters) as a function of cosmic time has been
compared with the corresponding predictions for the standard \lcdm
cosmology, with the aim of identifying observational tests that are
able to disentangle between the different cosmologies based on future
surveys (like the EUCLID satellite, \citealt{Laureijs11}).

In contrast, the influence of modified DE models on the properties and
the evolution of galaxy populations has not yet been explored in full
detail, even though many cosmological tests for constraining the DE
equation of state ultimately rely on a precise understanding of how
galaxies trace mass. This can in part be understood as a result of our
limited present understanding of galaxy formation even in the
$\Lambda$CDM cosmology, which is hampered by several long-standing
discrepancies between the predictions of theoretical models and
observations, as seen, for example, in the redshift evolution of the
galaxy stellar mass function (see e.g.~\citealt{Fontanot09b}), the
properties of the Milky Way satellites (see e.g.~\citealt{Maccio10,
  BoylanKolchin12}), the low baryon fraction in galaxy clusters (see
e.g.~\citealt{McCarthy07}), or in the shallow DM profiles associated
with different galaxy populations (``cusp-core'' problem, see
e.g.~\citealt{Moore94}). This in turn fuels some interest in exploring
both the possibility that these discrepancies may be reduced by
assuming an evolving DE, and in finding observational tests based on
statistical galaxy properties that could potentially distinguish
between different DE scenarios (which one may also hope to be easier
to perform compared with tests working purely in the ``dark sector'').

The catch of course is that galaxy evolution is a complex process,
involving a non-linear blend of many different physical processes
acting on the baryonic gas. All models of galaxy formation on
cosmological scales (both semi-analytic models, SAMs hereafter, and
numerical simulations alike) need to simplify this intrinsic
complexity by means of quite coarse, yet physically grounded, analytic
approximations (in SAMs) or sub-grid models (in simulations). In SAMs,
these analytic approximations are meant to describe the physical
processes acting on the baryonic gas (such as gas cooling, star
formation and feedback) as a function of the physical properties of
model galaxies (like their cold gas and stellar content). This then
yields a convenient parametrisation of the physics, which is
calibrated against a small subset of low-redshift observations. The
ability of modern models of this kind to successfully match a large
and diverse set of observations not used in the calibration can be
viewed as a powerful confirmation of the viability of the basic
paradigm of hierarchical galaxy formation.

However, it has also been shown that the SAM approach entails a
significant level of degeneracies among the different parameters
\citep[e.g.][]{Henriques09}, and the theoretical uncertainty is
exacerbated by the fact that different models tend to adopt different
parametrizations for the physical processes acting on the baryonic
gas. Even though the comparison of different model predictions shows a
reassuring level of consistency in many cases (see
e.g. \citealt{Fontanot09b}), there is hence a certain degree of
systematic uncertainty in this approach.  Therefore, in order to
assess the role of modified DE on galaxy evolution and, viceversa, to
use galaxy formation to constrain DE models, it is important to not
only quantify the impact of modified cosmologies on galaxy formation
within a specific model, but also to check how the size of the changes
in the model predictions compares to the intra-model variance for a
fixed $\Lambda$CDM cosmology.

This paper is organized as follows. In Section~\ref{sec:models}, we
introduce the cosmological numerical simulations and semi-analytic
models we use in our analysis. We then compare the various predictions
in Section~\ref{sec:results}. Finally, we discuss our conclusions in
Section~\ref{sec:final}.

%%%%%%%%%%%%%%%%%%%%%%%%%%%%%%%%%%%%%%%%%%%%%%%%%%%%%%%%%%%%%%%%%%%%%%%%%%%%%%
\section{Models}\label{sec:models}
\begin{figure*}
  \centerline{ \includegraphics[width=18cm]{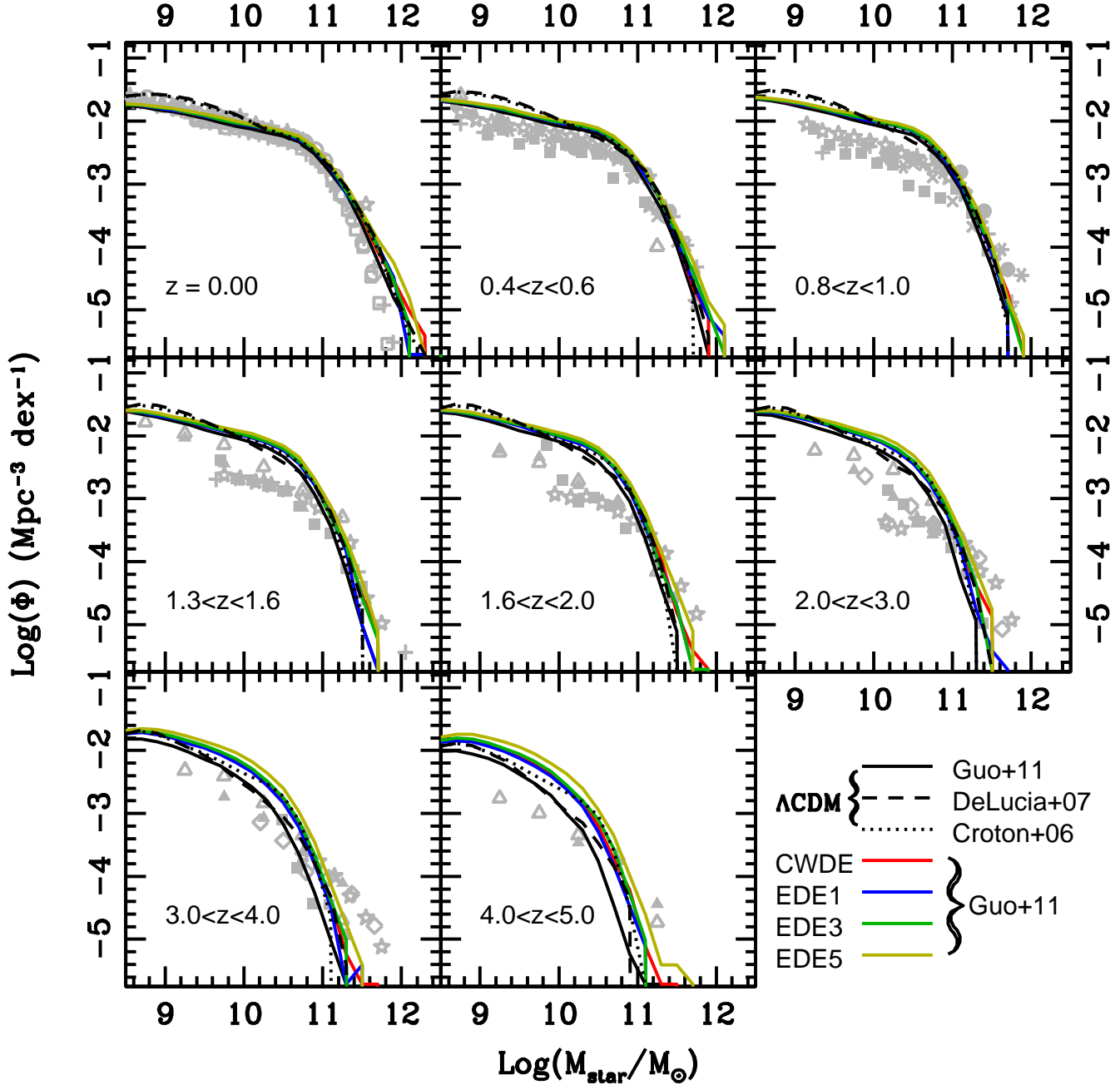} }
  \caption{Redshift evolution of the galaxy stellar mass function in
    different cosmological models, and for different SAMs in the same
    $\Lambda$CDM model. Various line styles and colours distinguish
    the different models, as labeled in the legend.  Grey points refer
    to the stellar mass function compilation from \citet[][see
      references herein]{Fontanot09b}.}\label{fig:gsmf_evo}
\end{figure*}
\begin{table}
  \caption{Cosmological Parameters and Model definition.}
  \label{tab:cosmo}
  \renewcommand{\footnoterule}{} \centering
  \begin{tabular}{ccccccc}
    \hline
     & $\Omega_m$ & $\Omega_\Lambda,0$ & $h$ & $\sigma_8$ & $w_0$ & $\Omega_{\rm de,e}$ \\
    \hline
    \lcdm & 0.25 & 0.75 & 0.73 & 0.9 & -1.0  & --- \\
    CWDE  & 0.25 & 0.75 & 0.73 & 0.9 & -0.6  & --- \\
    EDE1  & 0.25 & 0.75 & 0.73 & 0.9 & -0.93 & 2 $\times 10^{-4}$ \\
    %EDE2  & 0.25 & 0.75 & 0.73 & 0.9 & -0.99 & 8 $\times 10^{-4}$ \\
    EDE3  & 0.25 & 0.75 & 0.73 & 0.9 & -0.93 & 2 $\times 10^{-3}$ \\
    %EDE4  & 0.25 & 0.75 & 0.73 & 0.9 & -0.99 & 8 $\times 10^{-3}$ \\
    EDE5  & 0.25 & 0.75 & 0.73 & 0.9 & -0.93 & 2 $\times 10^{-2}$ \\
    \hline
  \end{tabular}
\end{table}

\subsection{Dark energy parametrisation}

In this paper, we focus on a specific class of DE cosmologies, called
{\it Early Dark Energy} (EDE) models. In these cosmologies, DE
constitutes an observable fraction of the total energy density at the
time of matter-radiation equality (in contrast to the cosmological
constant scenario). A natural outcome of these models is to predict an
earlier formation of structures with respect to the \lcdm cosmology
for an equal amplitude of the present-day clustering strength. In
fact, in EDE models the high-$z$ cluster population grows considerably
relative to $\Lambda$CDM, helped also by a lowered value of the
critical linear density contrast needed for collapse
\citep{Bartelmann06}; as a result, EDE models predict a slower
evolution of the halo population than the standard \lcdm cosmology.

In particular, we consider similar EDE models as in
\citet{GrossiSpringel09}, as originally introduced by
\citet{Wetterich04}. These models are characterized by a low but
non-vanishing dark energy density at early times, and by a low
redshift value of $\Omega_\Lambda$ consistent with CMB
constraints. The equation of state parameter $w(z)$ varies with
time according to the equation
\begin{equation}\label{eq:ede1}
w(z) = \frac{w_0}{1+b \ln^2(1+z)} ,
\end{equation}
\noindent
where
\begin{equation}\label{eq:ede2}
b = - \frac{3\, w_0}{\ln \left( \frac{1-\Omega_{\rm
      de,e}}{\Omega_{\rm de,e}} \right) + \ln \left(
  \frac{1-\Omega_{\rm m,0}}{\Omega_{\rm m,0}} \right )} .
\end{equation}
\noindent
In the previous equations, $w_0$ and $\Omega_{\rm de,0} =
1-\Omega_{\rm m,0}$ represent the present-day equation of state
parameter and amount of dark energy, respectively, while $\Omega_{\rm
  de,e}$ gives the average energy density parameter at early
times. For sufficiently low $\Omega_{\rm de,e}$, EDE models reproduce
the accelerated cosmic expansion observed in the local Universe.

We examine a set of different EDE models by changing $w_0$ and
$\Omega_{\rm de,e}$ (Table~\ref{tab:cosmo}). In particular, we vary
the $\Omega_{\rm de,e}$ parameter over two orders of magnitude; it is
worth stressing, however, that in our EDE5 model we assume a value for
$\Omega_{\rm de,e}$ which is rather extreme for this class of models
\citep{Hollenstein09}.

\subsection{Numerical simulations}

We have performed a series of numerical $N$-body simulations of
DM-only runs for all our EDE models. For comparison, we also run a
simulation in a standard \lcdm cosmology configuration and we consider
a model with constant equation of state $w=-0.6$ (CWDE) as well. In
all simulations, we assume a flat universe with matter density
parameter $\Omega_m=0.25$, Hubble parameter $h=0.73$, and Gaussian
density fluctuations with a scale-invariant primordial power spectrum
with spectral index $n=1$. The normalization of the linearly
extrapolated $z=0$ power spectrum is taken to be $\sigma_8=0.9$ for
all simulations.

We generate initial condition for all the cosmological models using
the {\sc n-genic} code and have run the simulations using the
cosmological code {\sc gadget-3} \citep[last described
  in][]{Springel05c}. For both codes we employ the versions modified
by \citet{GrossiSpringel09} to account for EDE cosmologies with a
redshift-dependent equation of state. All simulations have been run
using $432^3$ particles in periodic boxes $100\,h^{-1}{\rm Mpc}$ on a
side, corresponding to a mass resolution of $1.18 \times 10^9\, h^{-1}
M_\odot$. For each box, 64 simulation snapshots were saved, at the
same redshifts used in the Millennium project \citep{Springel05},
facilitating a straightforward comparison. Group catalogues have been
constructed using the friend-of-friend (FOF) algorithm with a linking
length of 0.2 in units of the mean particle separation. Each group has
then been decomposed into gravitationally bound substructures using
{\sc subfind} \citep{Springel01}. The resulting subhaloes are then
used to construct merger history trees as explained in detail in
\citet{Springel05}. Only subhaloes that retain at least 20 bound
particles after the gravitational unbinding procedure are kept for the
tree construction, thus implying a subhalo detection threshold of
$2.36 \times 10^{10}\, h^{-1} M_\odot$.

As a consistency check, we have measured the matter power spectrum of
the different simulation models at various redshifts, finding
consistent results with respect to \citet[][their
  Fig.~5]{GrossiSpringel09}. Similarly, our results for the halo mass
functions reproduce their findings.

\subsection{Semi-analytic models}
\begin{figure}
  \centerline{ \includegraphics[width=9cm]{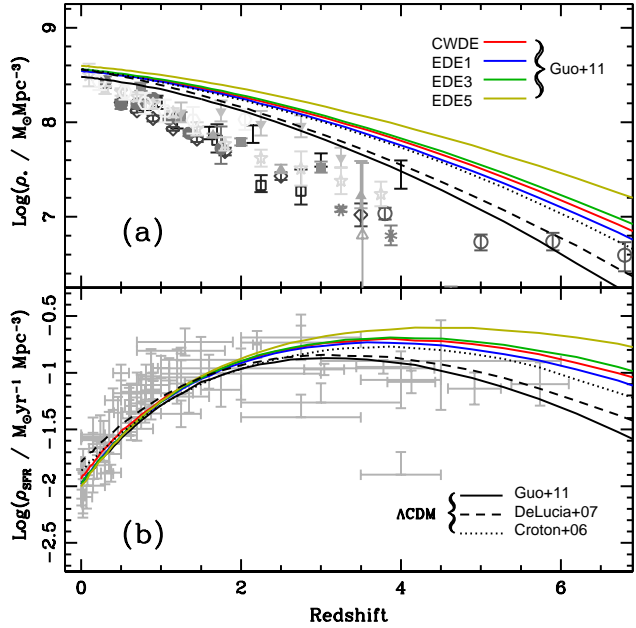} }
  \caption{Redshift evolution of (a) the stellar mass density, and (b)
    the cosmic star formation rate density. Grey points refer to a
    stellar mass density compilation from \citet[][see references
      herein]{Santini12} and to a cosmic star formation density
    compilation from \citet[][see references herein]{Hopkins04},
    respectively. Our models are shown with the same lines types and
    colours as in Fig.~\ref{fig:gsmf_evo}, as
    labeled.}\label{fig:csfr_evo}
\end{figure}

In this paper, we mainly consider the most recent implementation of
the {\it Munich} model \citep{Guo11}, based on the \munich code
originally developed by \citet{Springel05}. We also consider two
previous versions of the same code, namely the \citet{Croton06} and
\citet{DeLuciaBlaizot07} models. The use of different SAMs implemented
on top of the same underlying code structure allows an assessment of
the intra-model variance induced by different choices for the
approximation of galaxy formation physics.  This yields a course
estimate of systematic model uncertainties, which can in turn be
compared to the size of changes in the predicted galaxy properties due
to EDE cosmologies.  The chosen set of SAM models is particularly
suitable for this approach as it represents a coherent set of models
designed to work on Millennium-like merger trees, such that
complications such as merger tree conversions can be avoided.

\begin{figure}
  \includegraphics[width=9cm]{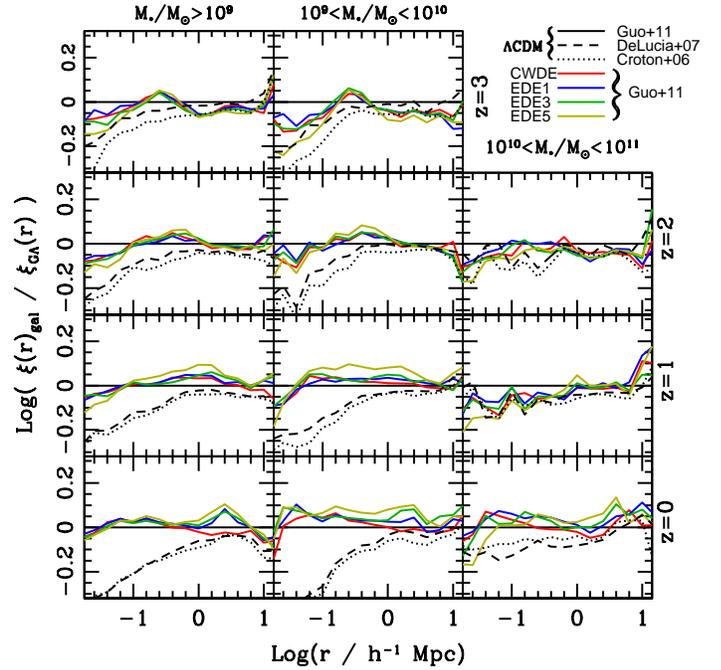} 
  \caption{Deviation of the galaxy auto-correlation functions in our
    model cosmologies with respect to the \citet{Guo11} model applied
    to the \lcdm simulation ($\xi_{\rm G\Lambda}$). The lines types
    and colours are the same as in
    Fig.~\ref{fig:gsmf_evo}.}\label{fig:tpcf}
\end{figure}

In the interest of conciseness, we refrain in the following from
discussing all the details of the modeling of different physical
processes in each of the SAMs (we refer the reader to the original
papers for a full specification). However, we want to point out where
the main differences between the three models lie: (a) in the
treatment of dynamical friction and merger times, the stellar initial
mass function and the dust model (from \citealt{Croton06} to
\citealt{DeLuciaBlaizot07}); (b) in the modeling of supernovae
feedback, the treatment of satellite galaxy evolution, tidal stripping
and mergers (improved from \citealt{DeLuciaBlaizot07} to
\citealt{Guo11}). In all cases, these major changes in the physical
recipes involved a re-calibration of the most relevant model
parameters.

\begin{figure*}
  \centerline{ \includegraphics[width=18cm]{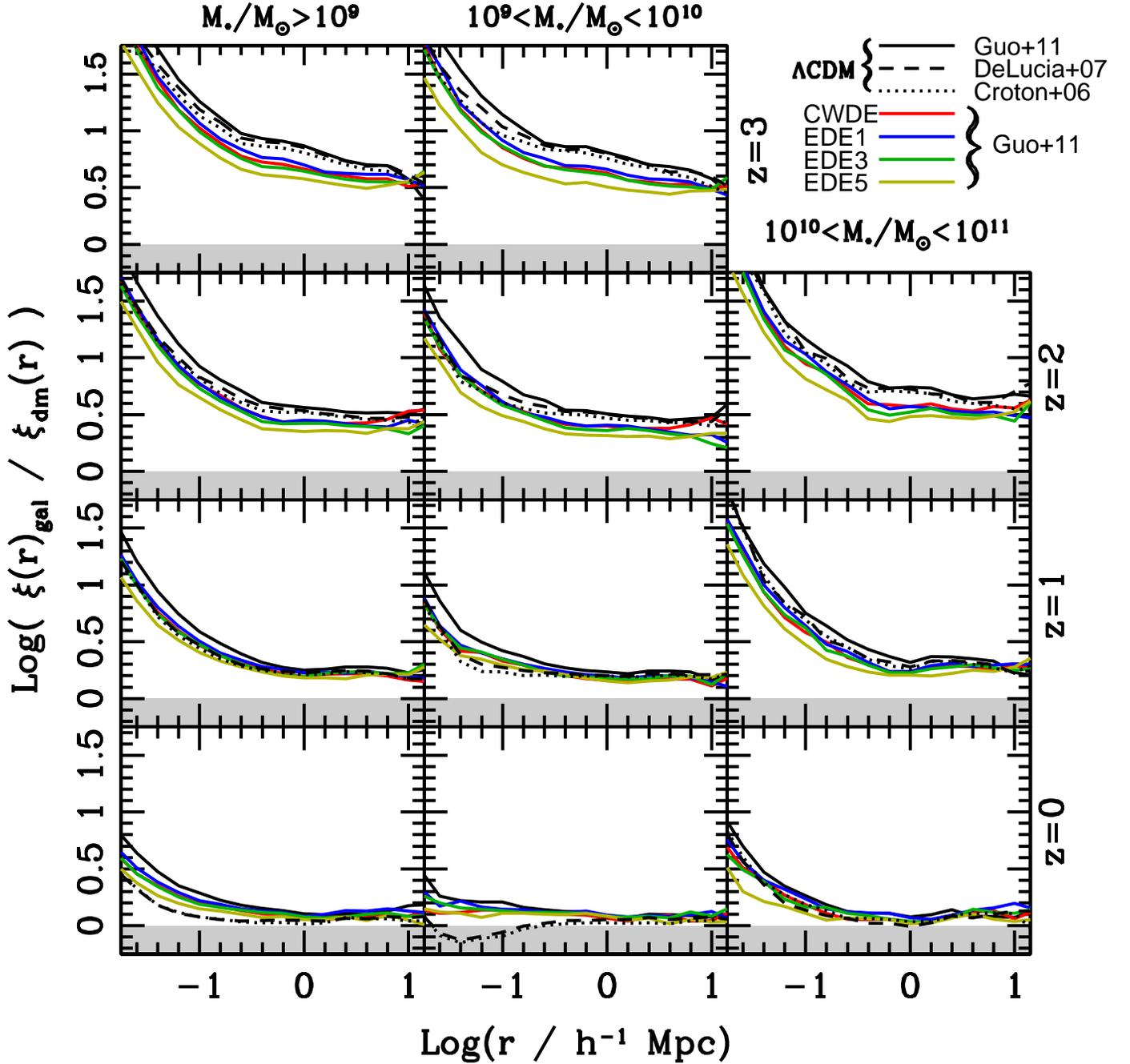} }
  \caption{Redshift evolution of the galaxy bias in different
    cosmological models. The lines types and colours are as in
    Fig.~\ref{fig:gsmf_evo}.}\label{fig:bias}
\end{figure*}

We note that we have modified the code used in the \citet{Guo11} model
in order to have it run self-consistently on EDE cosmologies, by
including the EDE contribution in the expression for the Hubble
expansion rate:
\begin{equation}
  H(a) = H_0 \sqrt{\frac{\Omega_{\rm m,0}}{a^3}+\Omega_{\rm de,0} \,
    e^{-3 \int [1+w(a)] d \ln a}}.
\end{equation}
This equation is a generalisation of the usual Hubble function
definition employed in \munich, and it naturally simplifies to the
correct forms when dealing with the CWDE and \lcdm cosmologies
($w(a)=-1$). In the following, whenever we discuss results based on
the \citet{Guo11} model, we actually refer to this modified code.

All SAMs have been calibrated by requiring them to reproduce a
well-defined set of low-redshift reference observations. In this
paper, we do not discuss possible changes in the model calibrations,
rather we prefer to retain the original parameter choices, in order to
allow a direct comparison to the published models and to highlight
differences induced by changes in the cosmology alone, for physics
models that are kept fixed. As far as the EDE and CWDE cosmologies are
concerned, this implies that these models are not necessarily tuned to
perform best, as in the $\Lambda$CDM case. Nonetheless, since we
require our numerical simulations to provide consistent statistical
properties for $z=0$ DM haloes by construction, we expect the
variations between SAM predictions in the different cosmologies to be
small at the lowest redshifts. In any case, our choice of keeping the
parameter choices used in the \citet{Guo11} model the same also in
different dark energy cosmologies is instrumental for testing the
effective change in model predictions due to a variation of the
assumed Hubble function {\it alone}, a distinctive feature of EDE
models.

%%%%%%%%%%%%%%%%%%%%%%%%%%%%%%%%%%%%%%%%%%%%%%%%%%%%%%%%%%%%%%%%%%%%%%%%%%%%%%
\section{Results \& Discussion}\label{sec:results}

We begin by comparing the redshift evolution of some of the most basic
global properties of the galaxy populations predicted by our SAMs,
namely the galaxy stellar mass function (Fig.~\ref{fig:gsmf_evo}), and
the evolution of the cosmic star formation rate and stellar mass
density (Fig.~\ref{fig:csfr_evo}).

In all the figures, the black lines show the different \munich models
in a \lcdm cosmology. In particular, the solid lines refer to the
\citet{Guo11} implementation of the model, while dashed and dotted
lines refer to the \citet{DeLuciaBlaizot07} and \citet{Croton06}
versions, respectively. The differences between the predictions of the
different SAMs applied to the same cosmological simulations are mainly
driven by the different calibration sets employed by the different
authors. For example, the \citet{Guo11} model has been especially
designed to reproduce the low-mass end of the stellar mass function at
$z=0$. Coloured lines, on the other hand, give the predictions of the
\citet{Guo11} model in different cosmologies. Red lines refer to the
CWDE simulation, whereas blue, green and yellow lines refer to the
different EDE realizations, as labelled in the legend.

We point out that the agreement of the \citet{Guo11} predictions among
the different cosmologies is satisfactory at $z=0$. When a comparison
to observational constraints is made, the model predictions have been
convolved with an adequate estimate of the typical observational error
(e.g.~a lognormal error distribution with amplitude 0.25 and 0.3 for
stellar masses and star formation rates respectively, see
\citealt{Fontanot09b}). The predictions of the \citet{Croton06} model
have been converted from a Salpeter to a Chabrier IMF by assuming a
constant shift of 0.25 dex in stellar mass and 0.176 dex in star
formation rate.

As expected, the earlier structure formation epoch in EDE cosmologies
directly translates into an increase of the space density of galaxies
at all redshifts (Fig.~\ref{fig:gsmf_evo}). The strength of this
effect and the redshift range involved increase with the $\Omega_{\rm
  de,e}$ value, as then the dark energy contribution to the total
energy density of the Universe becomes more relevant.  Nonetheless, it
is worth noting that this increase in space density is not limited to
the massive end of the stellar mass function, but is present at all
mass scales. The well-known problem of an overprediction of
intermediate-to-low mass galaxies in theoretical models of galaxy
formation \citep{Fontanot09b} is thus not reduced, but even enhanced
in EDE cosmologies. Following this argument, we also expect that a
number of known tensions between model predictions and the properties
of galaxies in the Milky-Way environment should not be solved but
exacerbated, although dedicated higher resolution simulations are
needed to firm up this conclusion. However, when we consider the
predictions of the other two \munich versions based on the same \lcdm
cosmology, we find a small spread in the predictions of different SAM
models computed for the same cosmology. This ``systematic
uncertainty'' in the theoretical modelling is of the same order as the
size of the differences seen in the predictions of a single SAM model
applied to different cosmologies.

Similar conclusions can be drawn by considering integrated quantities,
like the cosmic stellar mass density and mean star formation rate
density (Fig.~\ref{fig:csfr_evo}). In these quantities, the different
redshift evolution among different cosmologies and models is even more
clear: out to redshift $z\sim 4$, it is difficult to separate the
impact of different cosmologies from the effect of a different
modeling of physical processes and/or different model calibrations,
while at higher redshift, EDE models effectively predict somewhat
higher star formation rates and accumulated stellar mass density.
Anyhow, also in this case, only rather extreme models differ
significantly from the locus of considered \lcdm SAMs.

The galaxy auto-correlation function $\xi_{\rm gal}$ represents
another potentially useful discriminant between different
cosmologies. We consider auto-correlation functions corresponding to
different intervals of stellar mass. The predicted galaxy
auto-correlation functions for our models show good consistency among
each other over a wide redshift range, so they provide no direct way
to disentangle between different cosmologies. In Fig.~\ref{fig:tpcf},
we show the resulting auto-correlation functions normalised to the
predicted values for the \citet{Guo11} model in the \lcdm realization
($\xi_{G\Lambda}$), for all galaxies more massive than $10^9 M_\odot$
and in two different mass bins. Similar trends are seen in all
cosmologies, with no relevant deviations from the reference model in
most cases. Most notably, the strongest deviations are seen for the
\citet{Croton06} and \citet{DeLuciaBlaizot07} predictions at very
small scales. This is clearly due to the different treatment of
satellite galaxy evolution in \citet{Guo11}. For the theoretical
predictions on these small scales, the physical mechanisms acting on
baryonic gas dominate over the cosmological effects.

Nonetheless, these results can provide interesting insights when
coupled with additional information on the underlying distribution of
DM particles. In particular, for each cosmology, we have computed the
auto-correlation function $\xi_{\rm dm}$ of the mass for a randomly
selected subsample of DM particles (corresponding to $1 \%$ of the
total particles in the cosmological box, for computational convenience
-- a larger fraction would not change the results). As expected, the
redshift evolution of $\xi_{\rm dm}$ shows large deviations among
different cosmologies. We combine these complementary informations
into an estimate of the galaxy bias, defined as $\xi_{\rm gal} /
\xi_{\rm dm}$ (Fig.~\ref{fig:bias}). In all stellar mass intervals we
consider, it is in principle possible to distinguish between the
different cosmological models with the help of the bias, by
considering the intermediate to large scales (i.e.~larger than a few
tenths of a Mpc) at $z>0$. In fact, at these scales, there is a
concordance of the predictions for the bias from the different \lcdm
based SAMs, while, at the same time, the deviations of EDE models from
a standard \lcdm cosmology become progressively larger at increasing
redshift and for larger stellar mass. At smaller scales, the intrinsic
intra-model dispersion between SAM predictions due to the different
treatments of the satellite galaxy physics hampers firm cosmological
conclusions.

\section{Conclusions}\label{sec:final}
In this paper we discuss the expected impact of Early Dark Energy
(EDE) cosmologies on the properties of galaxy populations, as
predicted by semi-analytic models. We consider the latest version of
the \munich model \citep{Guo11}, modified to account for
time-dependent variations of the equation of state parameter $w$. We
also consider earlier versions of the same SAM \citep{Croton06,
  DeLuciaBlaizot07} to compare the size of differences in galactic
properties induced by changes in the underlying dark energy model with
the intra-model variance due to differences in the modeling of the
baryonic physics. The extension of our approach to consider other
cosmological models with alternative dark energy scenarios \citep[see
  e.g.][]{Baldi12} is in principle straightforward, provided
high-resolution N-body simulations of such scenarios can be
calculated. The latter has very recently become possible even for more
complicated theories of gravity, such as DGP or $f(R)$-gravity
\citep[see e.g.][]{Oyaizu08, Schmidt09b, KhouryWyman09,
  LiZhaoKoyama12}, such that our method can be fruitfully used to
connect such scenarios with observations of galaxies.  We plan to
expand our present results into this direction in forthcoming work.

Our results highlight that EDE cosmologies lead to important
modifications in the galaxy properties with respect to a standard
\lcdm universe. Nonetheless, they also show that these deviations are
of the same order of magnitude as those induced by different
assumptions in the modeling of physical properties and/or by parameter
recalibrations in a homologous set of $\Lambda$CDM-based SAMs
\citep[see also][]{Guo12}. Stronger effects are seen at higher
redshift ($z>4$), but these redshifts correspond to a cosmic epoch
where direct observational constraints on galaxies are extremely
difficult to obtain and where we expect SAM predictions to be
comparatively uncertain (given that their usual calibration samples
are selected at lower redshifts). This behaviour is mainly due to our
still limited understanding of the physical properties acting in the
baryonic sector, and to the high level of degeneracy between the
analytic approximations employed in different SAMs. We thus conclude
that predicted galaxy properties alone provide only weak constraints
on disentangling a standard \lcdm cosmology from an Early Dark Energy
model. These results are consistent with studies that analysed the
response of SAMs against variations of cosmological parameters
(corresponding to the best constraints from different data releases of
the WMAP satellite) in the \lcdm Universe \citep[see e.g][]{Wang08,
  Guo12}.

We have also shown that some of these degeneracies are broken if
additional information on the properties and distribution of the
underlying DM field became available. The combination of the stronger
evolutionary patterns expected for the ``dark sector'' and the
definition of galactic populations with suitably stable properties in
SAMs indeed leads to the identification of reliable cosmological
tests. In particular, we show the potential of a measurement of galaxy
bias on scales larger than a few tenths of Mpc as a cosmological
discriminant. Future observational efforts aimed at constraining $w$
and its redshift evolution, like the EUCLID mission
\citep{Laureijs11}, are indeed devised in order to provide, at the
same time, constraints on the DM distribution (via weak lensing) and
on the properties of large galaxy populations (via spectroscopy and
photometry). We stress however, that this mission is designed to
primarly focus on the $z\sim1$ diagnostics, where the predicted
effects of EDE cosmologies on galaxy bias are smaller than at higher
redshifts.

While we confirm that the combination of these different probes is
indeed one of the most promising methods to achieve constraints on the
dark energy component of our Universe, we also would like to emphasise
the power of the particular simulation approach taken here, which is
able to accurately predict galaxy bias in non-standard cosmologies as
a function of scale, epoch, and galaxy type. The combination of
high-resolution dark matter simulations and semi-analytic galaxy
formation models used in this paper should also be highly useful to
explore and understand generic effects of galaxy formation on
cosmological constraints, for instance those derived from BAO or
redshift space distortion measurements \citep{Angulo12}. This would
ultimately lead to an optimal and accurate exploitation of the next
generation of galaxy surveys.

\section*{Acknowledgements}
FF and VS acknowledge financial support from the Klaus Tschira
Foundation and the Deutsche Forschungsgemeinschaft through Transregio
33, ``The Dark Universe''. REA and BH are supported by Advanced Grant
246797 ``GALFORMOD'' from the European Research Council.

\bibliographystyle{mn2e}
\bibliography{fontanot}

\end{document}